\documentclass[draft]{agujournal2018}
\usepackage{apacite}
\usepackage{url} 

\usepackage{multirow}
\usepackage{makecell} 
\setcellgapes{5pt}

\usepackage{color}

\draftfalse

\journalname{JGR-Space Physics}
\begin{document}
\title{Automated classification of plasma regions using 3D particle energy distributions}



\authors{Vyacheslav Olshevsky\affil{1,2}, Yuri V. Khotyaintsev\affil{3}, Ahmad Lalti\affil{3,4}, Andrey Divin\affil{5}, Gian Luca Delzanno\affil{6}, Sven Anderz\'en\affil{1}, Pawel Herman\affil{1}, Steven W. D. Chien\affil{1}, Levon Avanov\affil{7}, Andrew P. Dimmock\affil{3}, Stefano Markidis\affil{1}}

\affiliation{1}{KTH Royal Institute of Technology, Stockholm, Sweden}
\affiliation{2}{Main Astronomical Observatory, Kiev, Ukraine}
\affiliation{3}{Swedish Institute of Space Physics, Uppsala, Sweden}
\affiliation{4}{Uppsala University, Uppsala, Sweden}
\affiliation{5}{St. Petersburg State University, St. Petersbourgh, Russia}
\affiliation{6}{Los Alamos National Laboratory, USA}
\affiliation{7}{NASA Goddard Space Flight Center, Greenbelt, USA}


\correspondingauthor{Vyacheslav Olshevsky}{sya@mao.kiev.ua}



\begin{keypoints}
\item We develop a technique for automated classification of plasma regions traversed by the MMS spacecraft
\item Our model distinguishes solar wind, megnetosheath, magnetosphere, and ion foreshock with $98$\% accuracy
\item It can be used to detect mixed plasma regions, bow shock and magnetopause crossings.
\end{keypoints}


\begin{abstract}
We investigate the properties of the ion sky maps produced by the Dual Ion Spectrometers (DIS) from the Fast Plasma Investigation (FPI).
We have trained a convolutional neural network classifier to predict four regions crossed by the MMS on the dayside magnetosphere: solar wind, ion foreshock, magnetosheath, and magnetopause using solely DIS spectrograms.
The accuracy of the classifier is $>98$\%.
We use the classifier to detect mixed plasma regions, in particular to find the bow shock regions.
A similar approach can be used to identify the magnetopause crossings and reveal regions prone to magnetic reconnection.
Data processing through the trained classifier is fast and efficient and thus can be used for classification for the whole MMS database.
\end{abstract}

\section{Introduction}
Over the recent decades, missions such as Cluster, THEMIS, and MMS, have provided the space physics community with an abundance of in situ measurements across the magnetosphere and the solar wind.
These regions contain internally distinct plasma and field characteristics, which correspond to important regions and boundaries (e.g bow shock, magnetopause, foreshock) that are of high scientific interest.
On many occasions, scientific investigations are centered explicitly on the physical processes operating at these regions/boundaries.
However, before that, they must be manually identified in the data.
The current state of available measurements encompasses decades of observations, and manually surveying these data and choosing regions of interest is labor-intensive and often ineffective.
The combination of an improvement in the sophistication of machine learning techniques and the more immediately available computational resources, afford a means to classify and sort massive quantities of data.
In this paper, we describe a machine learning methodology that can automatically identify separate plasma regions across the upstream solar wind and dayside magnetosphere using MMS data.

The principal objective of the Magnetospheric Multiscale Mission (MMS) \citep{Burch:2016} is to understand the physical processes and the fundamental sequence of events causing magnetic reconnection since it is the central driver of space weather events at Earth and a fundamental plasma process across diverse plasma environments.
However, MMS also contains an instrument suite that delivers plasma and field measurements at an unprecedented temporal resolution, and therefore these observations are also suitable for investigating additional regions and processes, such as the dynamics of the bow shock and the magnetosheath.
Consequently, the ability to automatically identify these key regions across the entire MMS catalog is important in facilitating such investigations.
Each MMS spacecraft uploads on the ground about $4$ gigabits of data every day \citep{Burch:2016,Fuselier:2016}.
Some data pre-selection is performed by scientists-in-the-loop (SITL) \citep{Fuselier:2016} who continuously scan the data and decide which burst mode intervals should be stored and transmitted to the ground.
However, it does not implement a complete classification of the various plasma boundaries and regions measured by MMS.
Our aim is not to automate and replace SITL, but to complement it by providing a means to distinguish between key dayside regions.

When the MMS perigee is located on the dayside, the spacecraft traverses three well-defined plasma environments: pristine solar wind (SW), magnetosheath (MSH), and the magnetosphere (MSP).
The boundary between the MSP and the MSH is called the magnetopause, while a bow shock separates the SW from the MSH.
The structure of the bow shock can deviate significantly depending on the angle between the shock surface normal and the upstream magnetic field.
The shock is quasi-perpendicular (quasi-parallel) if this angle is $> 45^\circ$ ($< 45^\circ$).
In contrast to the sharp boundary of the quasi-perpendicular bow-shock, the quasi-parallel shock is much more structured, and identification of the shock boundary is much more challenging.
In the quasi-parallel part of the bow-shock, the SW and the MSH are separated by an ion foreshock (IF), which is a region consisting of the solar wind ions and the shock reflected ions that are directed upstream.
In addition, boundaries between the various plasma regions are not fixed but are affected by the rapidly varying solar wind conditions.
The detection of such boundaries is essential for achieving the scientific goals of the MMS mission not only because dayside magnetic reconnection occurs at the magnetopause \citep{Fuselier:2017}, but because MMS can also be used to analyze additional regions of fundamental interest.

We aim to develop an end-to-end technique for the automated classification of MMS dayside observations with no time-dependence and minimal data pre-processing and interpretation.
We introduce a simple and robust classifier model which classifies measurements of the dual ion spectrometers (DIS) of the Fast Plasma Investigation (FPI) suite \citep{Pollock:2016}.
It distinguishes between four plasma regions (SW, MSH, MSP, IF) using solely the ion energy distributions.
The model utilizes one measurement at a time, hence, it does not use any time history of the data, as we will demonstrate later, its estimated accuracy is $98$\%.
We describe the data in section~\ref{sec:data}, introduce the convolutional neural network (CNN) model in section~\ref{sec:model}, and present our results in section~\ref{sec:results}.

\section{Related work}
MMS Instruments operate in two time resolution modes: low (survey) and high (burst).
The data, collected by the instruments, is stored on board for a limited period of time.
All survey data is transmitted to the ground, while only a fraction $\sim 1-2$\% of the burst data can be transmitted due to telemetry limitations \citep{Baker:2016}.
Therefore a major effort of the SITL is devoted to selecting the regions of interest, such as magnetopause crossings.
The SITL produces reports in which they mark the suspected plasma region (e.g., MSH or MSP) for each observation interval, and select the burst intervals to be transmitted.
Machine learning methods have been applied to MMS data, first of all, to automate the selection of the burst intervals of highest interest, such as magnetopause crossings. 

\citet{Piatt:2019arXiv} developed a technique for the automated detection of the magnetopause using a statistical hierarchical Bayesian mixture model based on the data obtained from measurements of several instruments on MMS.
This model uses $153,127$ data samples, taken along $14$ randomly selected orbits with $4.5$ sec interval during January -- February 2017, and split $80/20$\% into the train/validation subsets. Each data sample is represented by several hand-engineered features, such as plasma density or ion temperature.
It employs the time-dependence of the data and requires data pre-processing to estimate the likeliness of each measurement to be a magnetopause.
According to \citet{Piatt:2019arXiv}, this single-class model gives $31$\% true positive rate and $93$\% true negative rate.
Such accuracy makes this model suitable for cross-checking the SITL selections, but it is hardly useful to do region selection.

The study of \citet{Argall:etal:2020} extended the idea of using hand-crafted features, conventionally used by the SITL, to automate the process of magnetopause crossing detection. 
Their binary classification model, GLS, consists of two long-short term memory layers followed by the dropout layers and uses $121$ features computed from the survey data transmitted to the ground.
The model was trained on the data taken during January $2017$, divided $80/20$\% in train/test subsets, to predict the regions of interest.
Applied to the data between $19$ October $2019$ and $25$ March $2020$, the GLS model has selected $360$ regions, out of which $71$\% were also selected by the SITL.
Out of these, $49$\% were also classified as magnetopause by the SITL.
Hence, the GLS model outperforms the automated burst system, but its ability to substitute the SITL is unclear.

\citet{Breuillard:2020} parsed the SITL reports in order to extract the keywords identifying $10$ different plasma regions in both the dayside and nightside campaigns.
Their one-dimensional deep convolutional neural network, FCN, consumes $3$-min intervals, consisting of $40$ data points each. At each point, the data is represented by $12$ hand-engineered features, similar to the ones used in the previous works.
The dataset consists of $34,159$ samples split into training ($56$\%), test ($25$\%) and validation ($19$\%) subsets.
Although the model has overfitted, according to Figure~1 in \citet{Breuillard:2020}, it demonstrates a good performance on the test dataset: the estimated precision varies between $84$\% (MP) and $96$\% (MSP), and recall between $84$\% (MP) and $95$\% (IFS, MSP).

The main purpose of the SITL reports is selection of the burst intervals to be downloaded, not the accurate labeling of the data for neural network training. 
We did manual labeling of two full months of MMS1 data, and found that SITL classification of plasma regions is not always precise in the dayside campaign. The SITL labels cover rather extended time intervals and often mislabel intermittent regions with a different type of plasma (e.g., MSH regions in the SW).
The FCN model has a rather high capacity (more than $1$ million free parameters), and has a potential to perform better given more fine-grained ground truth data. 

The above described models employ hand-crafted features for their predictions.
On one hand, features such as plasma density, are more intuitive for humans. But on the other, the model, trained solely on those features, does not `see' all the information collected by the instrument.
The predictive capability of such model, thus, is limited by the amount of data provided by the hand-crafted features.
However, as the experience from adjacent fields such as neural language processing or computer vision suggests, deep networks are able to learn data representation themselves.
Hence, a properly designed model should be able to produce accurate plasma region classification given the raw measurements.
We prove this concept by training a simple convolutional neural network solely on the three-dimensional ion energy spectra produced by DIS.

\section{Data}
\label{sec:data}
The FPI DIS data has been obtained from the \citet{mms-sdc} and represents the level-two data product of FPI, namely the ion energy distribution.
Each sample is an $[32, 16, 32]$ array, where the dimensions correspond to 32 energy, 16 polar angle, and 32 azimuthal angle bins.
The spacecraft coordinate system is close to the Geocentric Solar Ecliptic (GSE) system.
Measurements are taken with a 4.5-sec interval and saved in NASA's Common Data Format (CDF) \citep{nasa-cdf}.
Each CDF file contains no more than 2 hours of observations, or $1600$ measurements.
Each measurement is uniquely identified by the index of the spacecraft and epoch.
The epoch is counted in nanoseconds since the standard J2000 moment.

\begin{figure}[h]
\centering
\includegraphics[width=0.98\columnwidth]{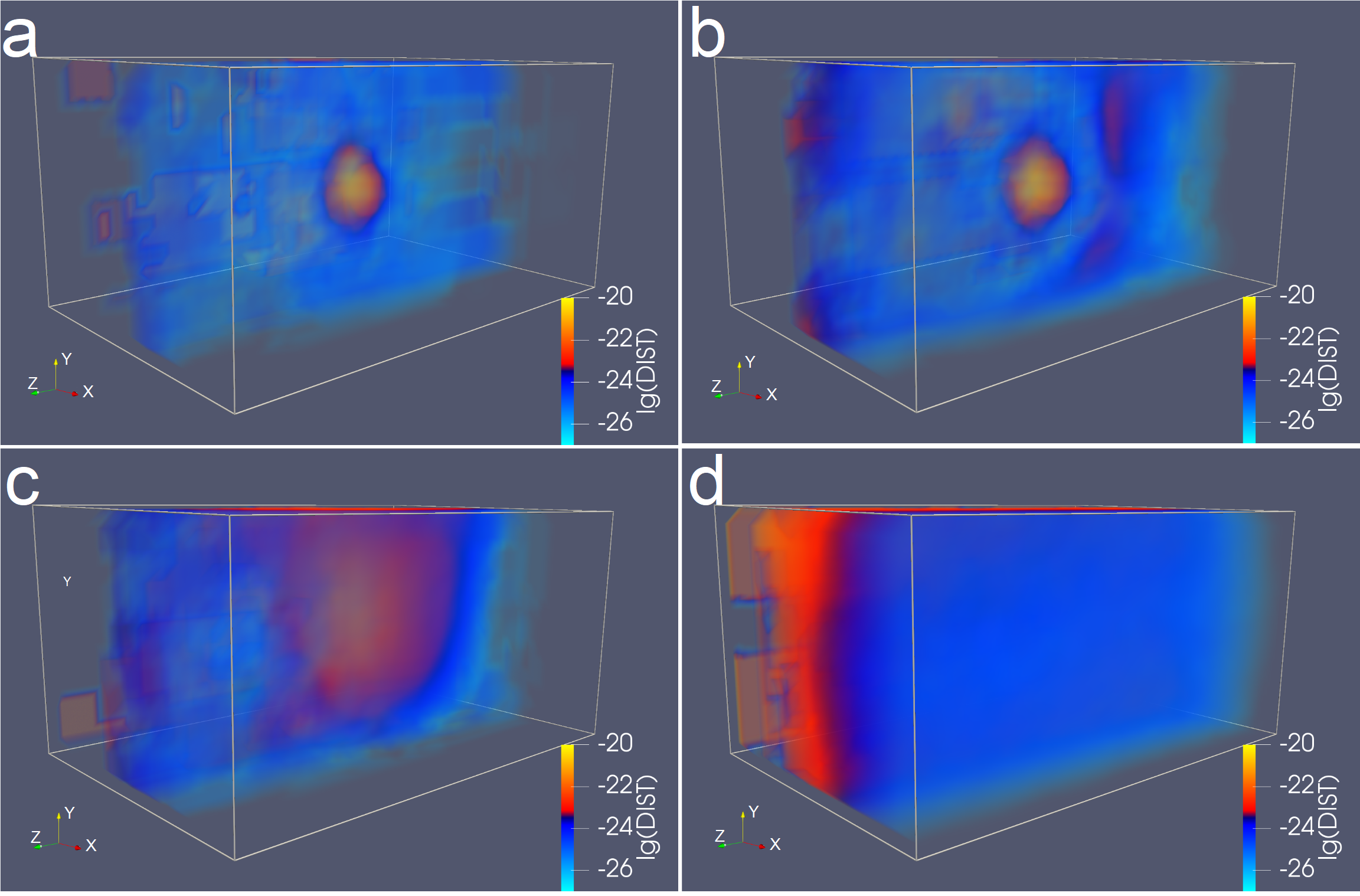}
\caption{Volume rendering of the logarithm of the ion phase space density for four typical observations.
In each panel, x corresponds to the energy bin, y to $\theta$, and z to $\phi$.
The axes are denoted in the lower left corner of each panel with red (x), yellow (y), green (z).
a) Solar Wind; b) Ion foreshock; c) Magnetosheath; d) Magnetosphere.}
\label{fig:dis3d}
\end{figure}

Typical samples of the ion distributions in the SW, IF, MSH, and MSP are shown in Figure~\ref{fig:dis3d}.
The data is visualized with a volume rendering of the logarithm of the ion phase-space density.
In this representation, each region exhibits some distinct characteristic features.
The SW is seen as an isolated sphere in the center of the box (Fig.~\ref{fig:dis3d}ab), corresponding to a narrow-energy ion beam.
The IF is a combination of the solar wind but with the manifestation of a ring feature surrounding it (Fig.~\ref{fig:dis3d}b).
This represents a mixture of the SW beam and the shock-reflected ions.
The MSH region in Fig.~\ref{fig:dis3d}c is apparent by the broadened distribution, which is formed by high-energy ions and the anisotropic velocity distribution.
The MSP is characterized by an isotropic distribution of high-energy ions and it can readily be distinguished from the other three classes by the lack of any prominent features (Fig.~\ref{fig:dis3d}d).
Therefore, based entirely on the features of the ion phase-space distribution, it is plausible to classify each of these regions.

\subsection{Human-labeled datasets}
\label{sec:dataset}
We have labeled $2$ sets of measurements taken by the DIS on MMS1 in 2017 where $278,110$ samples are taken in November, and $191,168$ in December, as described in Table~\ref{tab:dataset}.
To provide additional context during the classification we have also used the magnetic field measured by the Fluxgate Magnetometer (FGM) \citep{Russell:2016}.
Dedicated software has been developed to help interactively label intervals that belong to the same class.
Each sample has been assigned one of the following basic classes: $0$ -- Solar Wind; $1$ -- Ion foreshock; $2$ -- Magnetosheath; $3$ -- Magnetosphere and $-1$ -- Unknown.
Typical examples of different classes are shown in Figure~\ref{fig:dis3d}.
At the boundaries between the regions governed by these basic classes of distributions (such as the bow shock or the magnetopause), the distributions possess signatures of both classes of distributions on each of the sides of the boundary.
For example, at the magnetopause, the distribution is a mixture of the magnetosheath ($2$) and magnetospheric ($3$) distribution.
It is worth noting that the boundary layer distributions are not merely the weighted sums of the two adjacent classes, but they could be further complicated by other mixtures. For instance, the magnetopause region may contain plasma flows from magnetic reconnection.
As there is an infinite number of ways one can mix such distributions, we do not attempt to classify the mixed distributions and mark any kind of complex mixture or unclear situation as $-1$ -- Unknown.
The situations exhibiting mixed plasma populations can be of interest since they indicate the location of boundaries where interesting plasma processes are present, such as reconnection at the magnetopause.

\begin{table}[t]
 \caption{Human-labeled datasets} 
 \label{tab:dataset}
 \centering
 \begin{tabular}{r|c|l|r|r|r|r}
 \hline
        &  &       & \multicolumn{2}{c|}{November 2017} & \multicolumn{2}{c}{December 2017} \\
 \hline
  Label  & Token & Region & \# examples & percentage & \# examples & percentage  \\
 \hline
   $-1$  & UNK & Unknown & $41,685$ & $15.0$ & $28,332$ & $14.8$ \\
   $0$  & SW & Solar Wind & $119,934$ & $43.1$ & $69,719$ & $36.5$ \\
   $1$  & IF & Foreshock & $24,775$ & $8.9$ & $20,455$ & $10.7$  \\
   $2$  & MSH & Magnetosheath & $67,085$ & $24.1$ & $51,312$ & $26.8$ \\
   $3$  & MSP & Magnetosphere & $24,631$ & $8.9$ & $21,350$ & $11.2$ \\
 \hline
    &    & \multicolumn{1}{c|}{Total}         & $278,110$ &      &  $191,168$ &   \\
 \end{tabular}
\end{table}

\subsection{Data normalization}
\label{sec:normalization}
Each data sample (ion energy distribution) is a $[32,16,32]$ array of $32$-bit floating-point numbers, which correspond to the energy flux recorded at each angle/energy bin.
It is important to note that fluxes recorded in different energy channels and directions vary by orders of magnitude and for this reason, we utilize the logarithmic representation of the data.
As the array elements representing the energy/angle bins where no counts have been detected have $0$ (zero) value, we fill those elements with the minimum non-zero value in the array.
The x-axis of the detector coordinate system points towards the Sun and the solar wind beam appears at the edges of the box, i.e. it is divided between the smallest $\theta\approx 0$ and the largest $\theta \approx 2\pi$.
For convenience, we wrap each array along the $\theta$ dimension so that the SW beam appears in the center of the array.
 To summarize, the following procedures were applied to each data sample:
\begin{enumerate}
    \item Replace all zeros with the smallest non-zero value.
    \item Compute the logarithm $\log_{10}$.
    \item Normalize to values between $0$ and $1$ (subtract the minimum and divide by the maximum).
    \item Wrap along the third axis ($\theta$) by $16$ elements.
\end{enumerate}
Steps $1-4$ were applied for training the CNN classifier, while only steps $1-2$ were used for the preliminary analysis.

\subsection{Preliminary analysis}
To better understand the datasets available to us, we first performed Principal Component Analysis (PCA), which was developed in the last century by \citet{Pearson:1901:PCA} and, independently, by \citet{Hotelling:1933}.
The most comprehensive modern description of the method is given by \citet{Jolliffe:2002:PCA}.
PCA offers a computationally efficient method for multivariate analysis and visualization of high-dimensional data.
Given a set of $n$ measurements for (a large number of) $p$ random dependent variables, it finds Principal Component (PC) vectors of sizes $p$.
PCs then form an orthogonal basis for the data that serves to describe the variance in the data.
The first PC accounts for the largest possible variance, the second PC accounts for the second-largest possible variance, and so on.
Each PC is an eigenvector of the $p\times p$ covariance matrix and each sample (a point in $p$-dimensional space) can be projected onto the PCs.

We use the PCA method by scikit-learn \cite{scikit-learn}.
Given a 2-dimensional $n \times p$ array as an input, it computes the specified number of PCs and returns the reduced-dimensional data transformed into the basis of PC vectors, e.g., an array of $n\times 2$ if only $2$ first components are of interest.
This array contains the `projection' of each $n$-size measurement vector on the first $2$ PC basis vectors containing most variance of the dataset.

\begin{figure}[t]
\centering
\includegraphics[width=0.7\columnwidth]{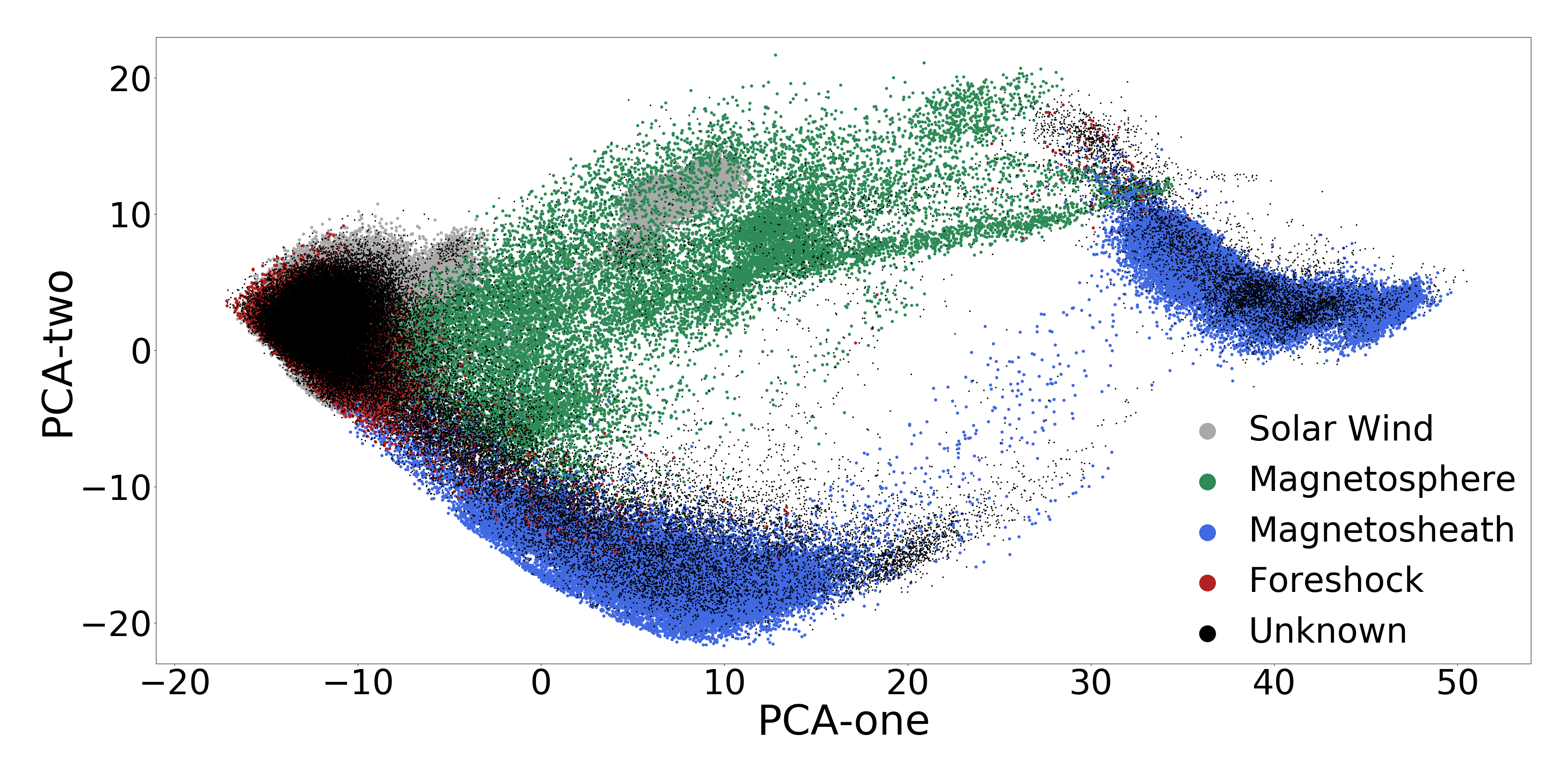}
\caption{Projection of the human-labeled 201711 dataset on the two first PCs.
Each circle represents one sample: SW (grey), MSP (green), MSH (blue), IF (red), UNK (black).}
\label{fig:PCA}
\end{figure}

We have computed the PCA over the human-labeled datasets (Table~\ref{tab:dataset}).
Only the steps $1$--$2$ of the data normalization (sec.~\ref{sec:normalization}) have been applied before computing the PCA.
It seems that the first two PCs represent approximately half of the explained variance, $40$\% and $9$\%, respectively.
We have plotted the $201711$ dataset against the first two PCs, to determine the extent to which the separate classes are distinguishable.
Examples of the same class tend to cluster in the diagram as shown in Fig.~\ref{fig:PCA}.
However, there is no clear separation between different clusters (`islands') in the plot.
The SW and IF clusters heavily overlap and there is no clear separation between the IF and MSH `islands'.
The first two PCs could not be applied to distinguish between different plasma regions, and $540$ PCs are required to cover $80$\% of the explained variance.
Even if those would allow for precise classification, an additional classifier is needed to attribute each sample to one of the four classes.

\begin{figure}[t]
\centering
\includegraphics[width=0.98\columnwidth]{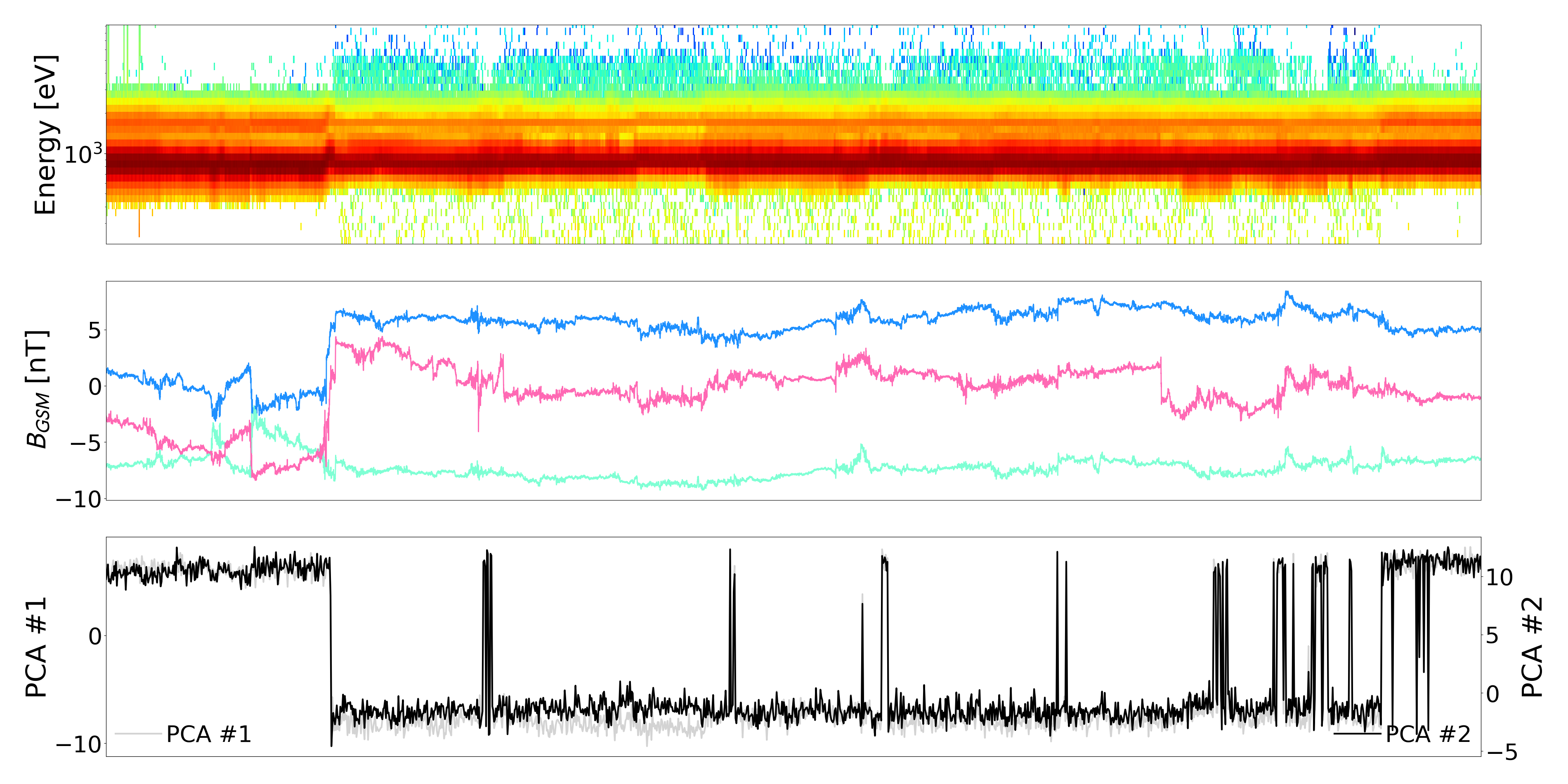}
\caption{Observations on 2017.11.15 12:00 -- 14:00.
a) $\log_{10}$ of the energy distribution integrated over $\phi$, $\theta$.
White pixels at high energy indicate no counts detected at higher energies in the colder solar wind which we replace with a very small value.
b) Blue, green, red: $B_{x,GSM}$, $B_{y,GSM}$, $B_{z,GSM}$.
c) Left axis, light grey: projection of the data on the first PC; right, axis, dark grey: projection of the data on the second PC.}
\label{fig:highenergy}
\end{figure}

\begin{figure}[h]
\centering
\includegraphics[width=0.9\columnwidth]{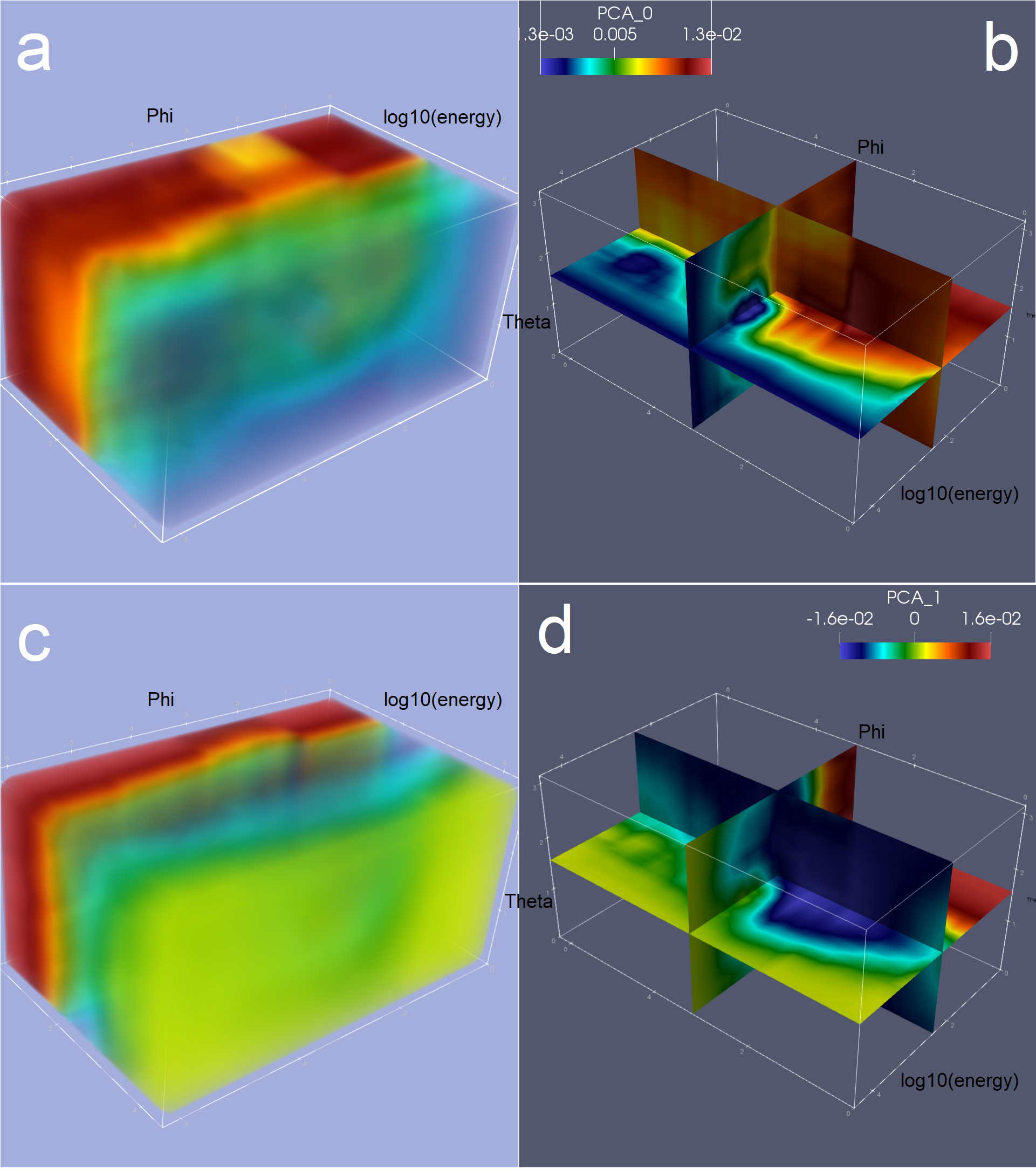}
\caption{a, b) Volume rendering and slice cuts through the first PC computed from the human-labeled dataset, reshaped to $32\times 16\times 32$ array.
Red color represents high variance, blue corresponds to low variance.
c,d) Same for the second PC.}
\label{fig:eigenvectors}
\end{figure}

Interestingly, we find that there are two SW clusters and two MSH clusters.
This splitting is not physical.
We found that this can be attributed to changes in the sensitivity of the detectors, which is changing dynamically depending on the total ion flux.
This is illustrated by an example in Figure~\ref{fig:highenergy}.
When the sensitivity is low, and no counts are detected far from the main peak of the distribution (at high and low energies), both PCA components are small.
When the sensitivity is high and there are counts detected at high and low energies, both PCA components are large.
This behavior could be further understood from the visualization of the first two PCs shown in Figure~\ref{fig:eigenvectors}, where the PC vectors are reshaped back to the shape of data samples ($32\times 16\times 32$).
The first PC is mostly sensitive to the counts detected at high energies and is nearly isotropic.
The second component is sensitive to the signals at the highest and the lowest energies.
Therefore the dimensionality reduction offered by the first two PCs omits very important variance in the center of the energy distribution array.
As we will show below, this is not a problem for the CNN classifier.

\section{Classifier model}
\label{sec:model}
\begin{figure}[t]
\centering
\includegraphics[width=0.3\columnwidth]{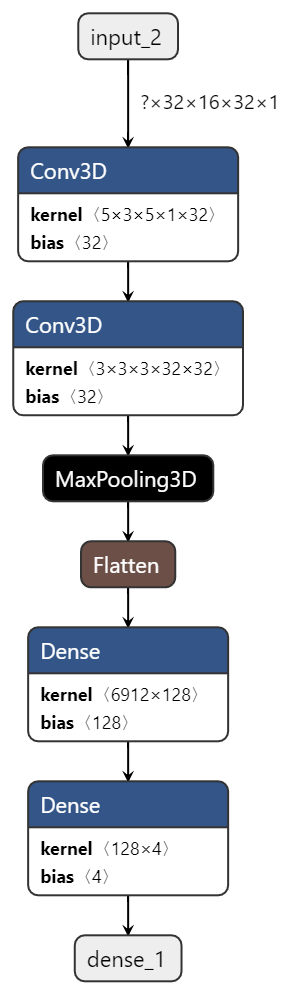}
\caption{The architecture of the classifier model.}
\label{fig:cnn}
\end{figure}

Impressive results in image classification have been achieved by CNNs recently \citep{Lecun:Yann:Bengio:2015}, however, there is no `universal' network architecture that is proven to be optimal across different tasks and datasets.
Taking into account the 3D nature of the data, it is natural to employ a 3D-CNN, rather than 2D-CNNs that are typically used for image classification.
We have developed a new 3D-CNN inspired by the VoxNet architecture proposed by \citet{Maturana:Scherer:2015} for object recognition in voxel arrays produced by robotic vision detectors.
It consists of 5 layers as illustrated in Fig.~\ref{fig:cnn}:
\begin{enumerate}
 \item 3D convolutional with (5, 3, 5) filters with strides (2, 1, 2) and no padding
 \item 3D convolutional with (3, 3, 3) filters with strides (1, 1, 1) and no padding
 \item 3D max-pooling
 \item Fully-connected layer with 128 elements
 \item Fully-connected layer with 4 elements
\end{enumerate}
The size of the filters and the strides were chosen to gradually decrease the spatial dimensions and increase the number of channels, enriching the representation in the feature space.
Given an input 3D array, the CNN outputs a set of probabilities for this sample to belong to each of the four classes: SW, IF, MSH, or MSP.
These probabilities sum to 1, and when one of the values is large, it could be considered a clean prediction (but not necessarily accurate).
While if two or more classes have rather high similar probabilities, most likely, the sample represents a region with mixed plasma populations, such as the bow shock or magnetopause.

\subsection{Training data}
It is impractical to use the whole dataset with $\sim200,000$ examples for training for the following reasons.
First, MMS spends a long time (tens of minutes) in plasma regions such as the SW or MSH, while it rather quickly traverses the IF.
As a result, both datasets are imbalanced (Table~\ref{tab:dataset}).
Second, energy distributions measured during traversing a uniform region, such as the SW, are very similar to each other, which was confirmed by a visual inspection of the data and by comparing the Euclidean distances between different examples.
Hence, there is little need to feed them all to a network being trained.
Lastly, the `Unknown' samples should be excluded from training.
Our experiments have shown that satisfactory results can be obtained when a rather small subset $\sim12\%$ of all samples is used for training.
The remaining samples could be employed for validation.

Several examples have been randomly chosen from each uniform (i.e., having the same label) plasma region from a human-labeled dataset.
Random selection coefficients were smaller for more abundant regions, SW and MSH, and higher for IF and MSP, to alleviate the class imbalance.
The random selection coefficients were chosen to select $\sim1/8$ of the total number of examples, excluding the UNK ones.
As a result, two well-balanced training datasets with $25,000$ samples were created, as listed in Table~\ref{tab:training}.

\begin{table}[t]
 \caption{Training datasets} 
 \label{tab:training}
 \centering
 \begin{tabular}{r|c|l|r|r|r|r}
 \hline
        &  &       & \multicolumn{2}{c|}{November 2017} & \multicolumn{2}{c}{December 2017} \\
 \hline
  Label  & Token & Region & \# samples & percentage & \# samples & percentage  \\
 \hline
   $0$  & SW & Solar Wind & $8,126$ & $31.8$ & $6,305$ & $24.9$ \\
   $1$  & IF & Foreshock & $5,954$ & $23.3$ & $6,601$ & $26.1$  \\
   $2$  & MSH & Magnetosheath & $5,306$ & $20.8$ & $5,612$ & $22.2$ \\
   $3$  & MSP & Magnetosphere & $6,146$ & $24.1$ & $6,757$ & $26.7$ \\
 \hline
    &    & \multicolumn{1}{c|}{Total}         & $25,532$ &      &  $25,275$ &   \\
 \end{tabular}
\end{table}

\subsection{Implementation}
We have trained two models: one on the 201711 training dataset, and one on the 201712 training dataset.
$20$\% of each training set was allocated for validation of the model's loss during training.
Each model is trained for $500$ epochs using an Adam optimizer \cite{kingma:2014} with learning rate $\alpha=1\cdot10^{-6}$, and a categorical cross-entropy loss
\begin{equation}
  L=-\sum_{i}^{C}t_i\log\left(f\left(s_i\right)\right),
  \label{eq:cross-entropy}
\end{equation}
where $t_i$ is the ground-truth probability (either $1$ or $0$), $C=4$ is the number of classes, and
\begin{equation}
  f\left(s_i\right)=\frac{e^{s_i}}{\sum_{j}^{C}e^{s_j}}
\end{equation}
is the predicted probability of the sample to belong to the $i^{th}$ class.
The implementation is done in Keras \cite{chollet2015keras} and TensorFlow~2 \cite{tensorflow2015-whitepaper}.
We used one NVIDIA GTX $1080$ GPU with $8$GB memory for training.
It takes a few hours to train our classifier on one training dataset (Table~\ref{tab:training}) for hundreds of epochs.
The datasets are saved and loaded in CDF files using the open-source cdflib module \citep{cdflib}.
The software which we developed for data pre-processing, classification, and visualization, is freely available on Bitbucket \citep{mmslearning}.
It works with the original folder hierarchy of the MMS data centers.

\section{Results}
\label{sec:results}
\begin{figure}[h]
\centering
\includegraphics[width=0.98\columnwidth]{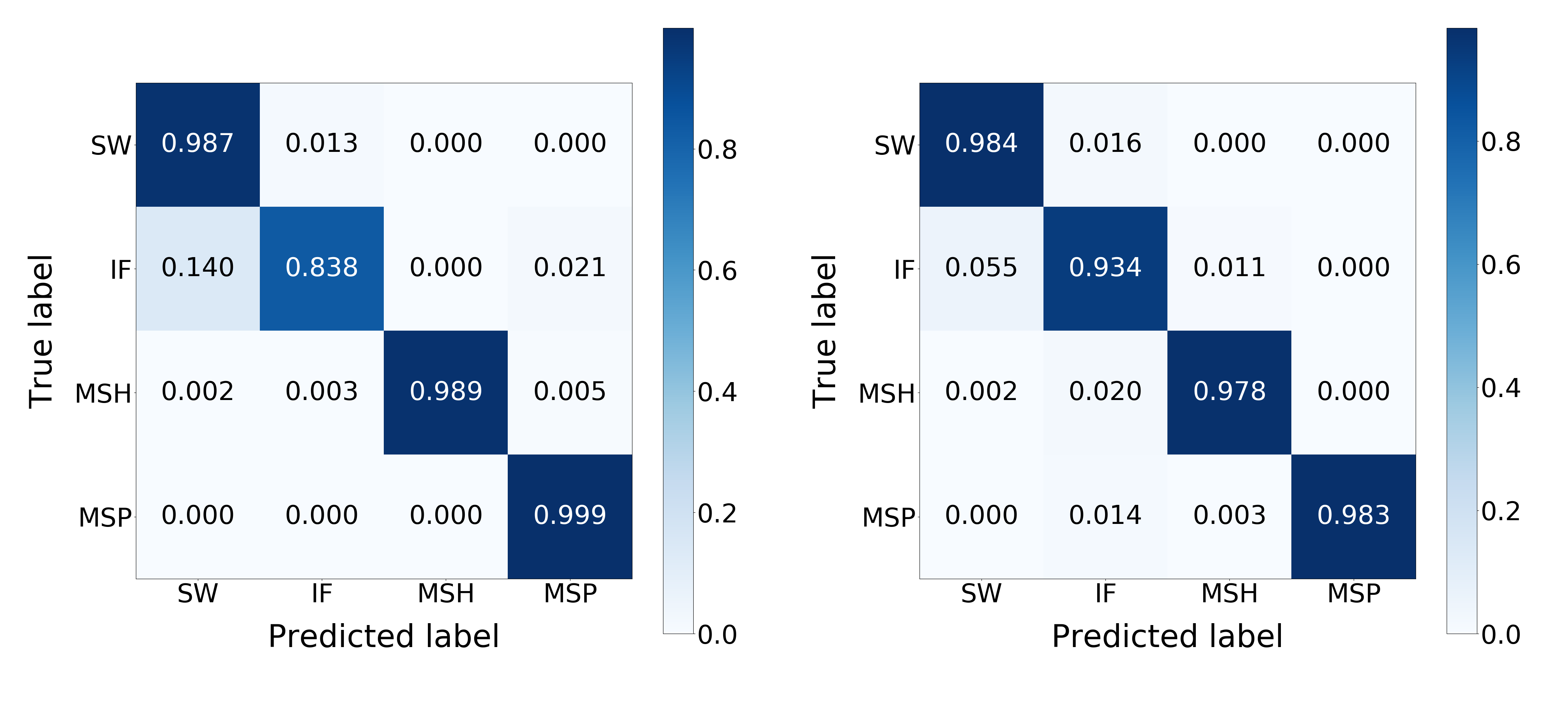}
\caption{Cross-validation confusion matrices.
Left: predictions of the model trained on 201711, for 201712 dataset.
Right: predictions of the model trained on 201712, for 201711 dataset.}
\label{fig:confusion}
\end{figure}

\begin{figure}[h]
\centering
\includegraphics[width=0.9\columnwidth]{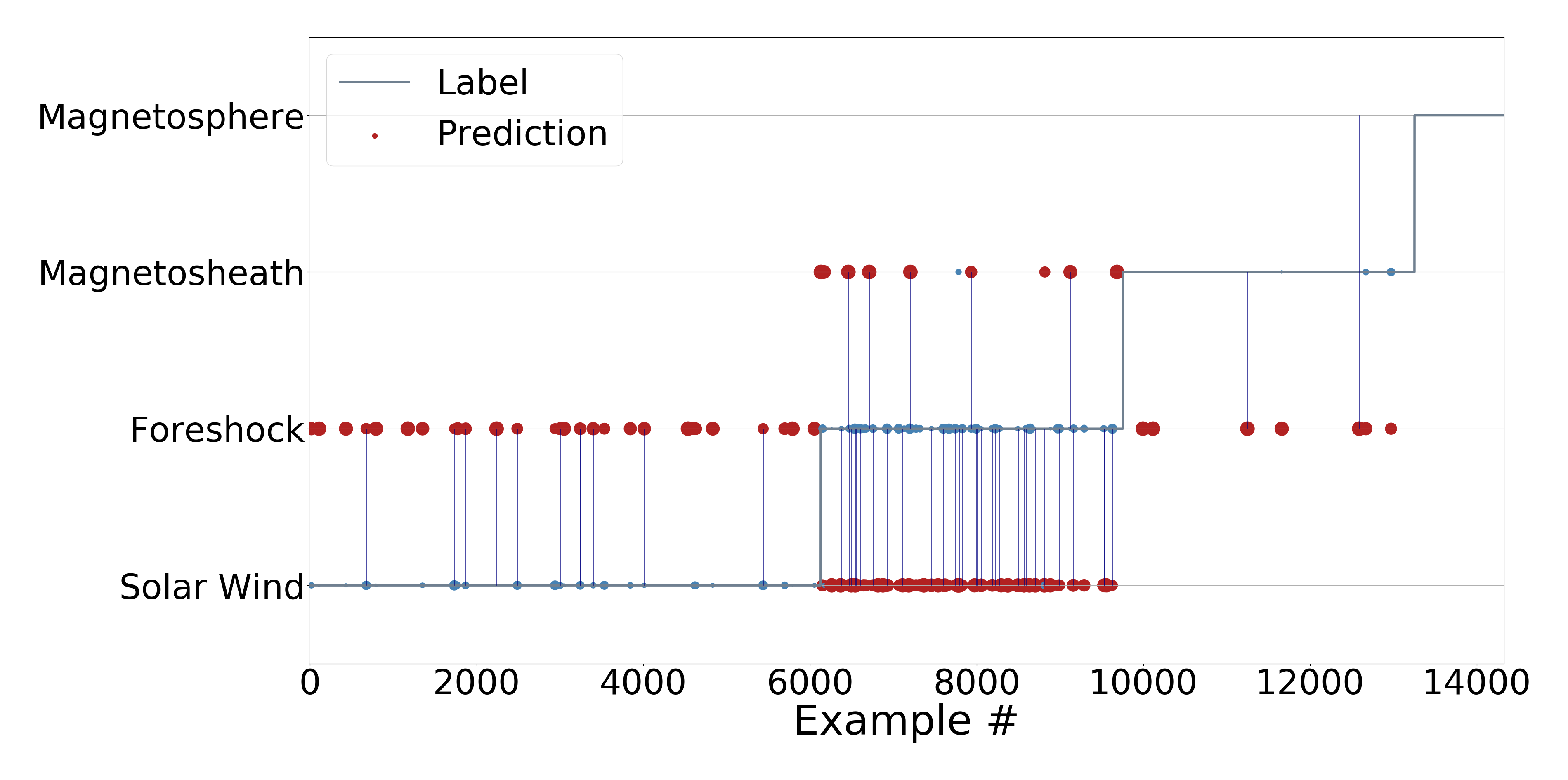}
\caption{A typical case in which the 201711-trained model misclassifies samples from the IF region.
Grey line is the reference human-made label.
Red circles correspond to the wrong predictions.
Blue circles represent the second most probable predictions for each mislabeled example.
The sizes of the circles are proportional to the predicted probabilities.}
\label{fig:evaluate}
\end{figure}

As noted above, we have trained two models on two different training datasets representing different months of observations.
Each of the models has been cross-validated against the human-labeled dataset for the other month (see Table~\ref{tab:dataset}), e.g. we have cross-validated the model trained on 201711 data against the 201712 dataset, and vise versa.
The 201711 model mislabels $1.61$\% of all known distributions from the 201712 dataset.
The 201712 model mislabels $1.09$\% of all 201711 distributions.
According to confusion matrices, based on the cross-validation (Figure~\ref{fig:confusion}), the most difficult to predict are distributions belonging to the IF class.
The accuracy (the ratio of the correctly classified samples to the total number of samples of this class) of the IF prediction is $83.8$\% for the 201711 model and $93.4$\% for the 201712 model.
Since the IF is a mixed plasma region, it is most often mislabeled as SW, and, in fewer cases, as MSP or MSH.
Figure~\ref{fig:evaluate} illustrates a typical case in which an IF region between the extended SW and MSH regions is mislabeled as either one of them.
In many cases, the second most probable prediction of the classifier is correct.
As the IF by definition contains the solar wind ions mixed with the shock-reflected ions and exhibit very low densities, such errors are to be expected, even in human-labeled datasets.
A possible approach for improvement might be to define the IF as a mixture of other classes.

\subsection{Mixed regions}
\begin{figure}[h]
\centering
\includegraphics[width=0.8\columnwidth]{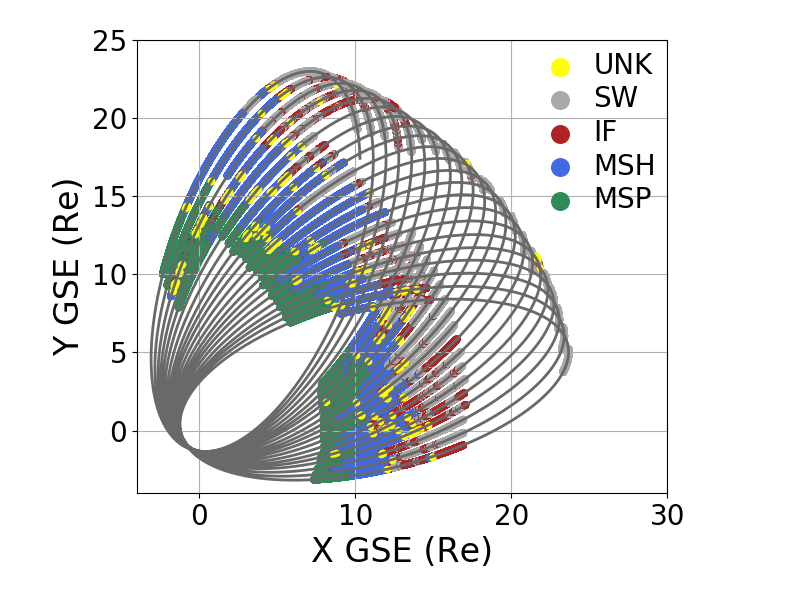}
\caption{Predictions of the trained CNN for two months of observations, 201711 and 201712 plotted in the $x_{GSE}$ and $y_{GSE}$ coordinates.
Color circles denote different regions, and the grey line shows the spacecraft trajectory.}
\label{fig:orbit:predictions}
\end{figure}

The model returns probabilities for each of the classes.
It appears that for $98.7$\% of samples from both human-labeled datasets, the top individual probability returned by each model is exceeding $0.7$.
On the other hand, the distributions for which the top score is below 0.5 contain mixing of different ion populations.
For example, Figure~\ref{fig:orbit:predictions} shows the predictions along the two-month trajectory where the distributions, for which the top scores are below 0.5 (`Unknown'), are marked with yellow circles.
These regions constitute $\approx0.04$\% of all the distributions, and most of such mixed states are located at the MSH/MSP and SW (IF)/MSH boundaries, i.e., they indicate the magnetopause and bow shock.

\subsection{Application: finding of the bow-shock}
To illustrate how an automated classifier can be used to aid scientific investigations, we apply the results of the model to identify bow-shock crossings.
We define a shock crossing as a sharp transition from either the SW/IF to the MSH, or vice versa.
To identify a crossing, we analyze the change in the probability provided by the model along the trajectory of the spacecraft.
Whenever the probability changes within a set period and the change is from the SW/IF to the MSH, we mark this time as a shock crossing.

\begin{figure}[h]
\centering
\includegraphics[width=0.9\columnwidth]{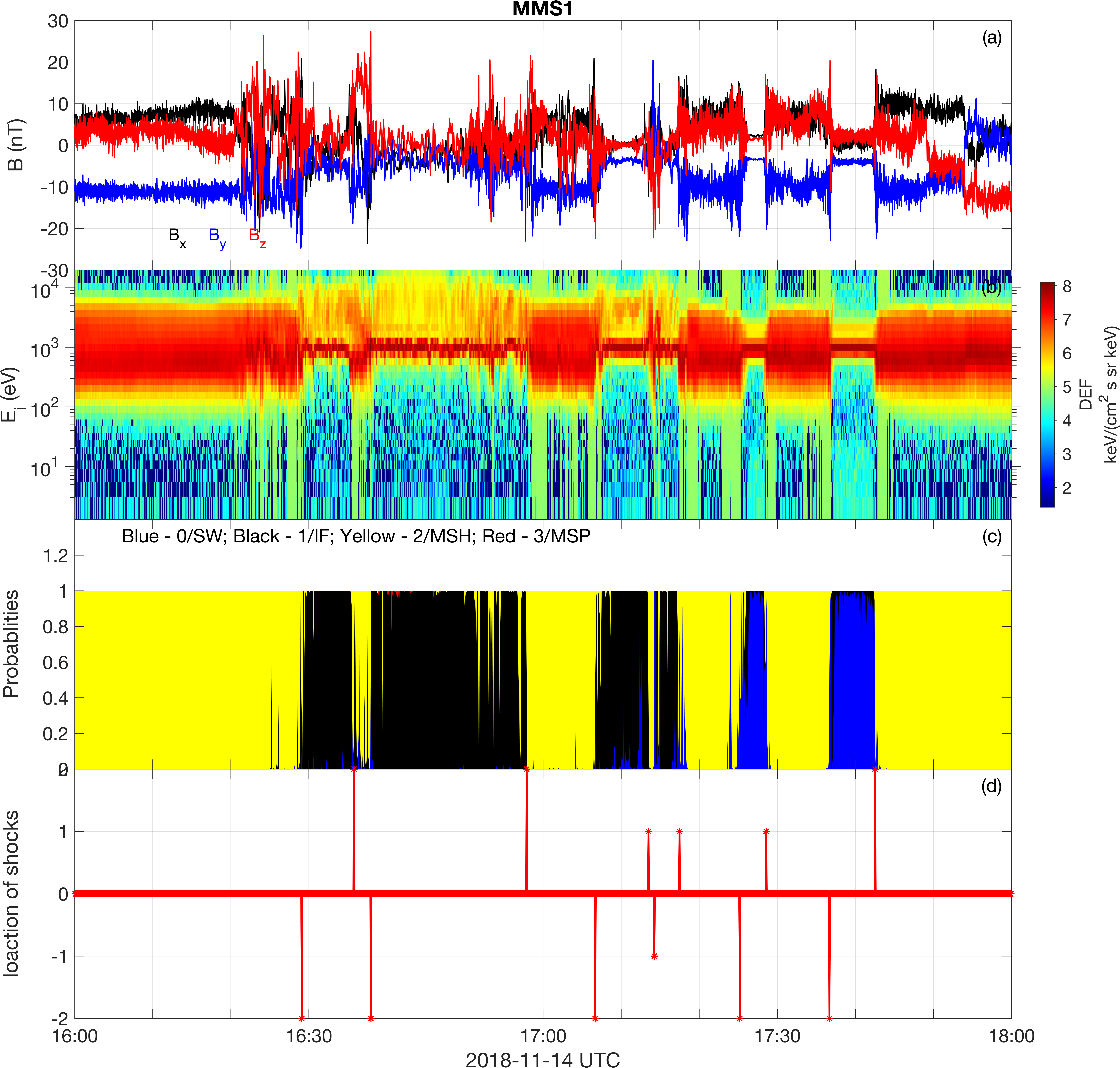}
\caption{Detection of shock crossing events using the probability output of the CNN model.
(a) magnetic field, (b) omni-directional ion differential energy flux (DEF), (c) probability output from the classifier, color coded with Blue being SW, black IF, yellow MSH and red MSP, (d) time location of shock crossings, marked by $1$ or $-1$, depending if it is a transition from upstream to downstream or vice versa.
These values are multiplied by 2 if the time corresponds to a burst interval from MMS.}
\label{fig:corrossings}
\end{figure}

An example interval where we performed this shock detection is shown in figure \ref{fig:corrossings}.
Panels (a) and (b) show the magnetic field and the omni-directional ion differential energy flux (DEF) respectively, from which one can visually identify that the spacecraft initially was in the MSH with a higher magnetic field and the broader DEF spectrum.
Then MMS crosses back and forth into the IF identified by the low amplitude turbulent magnetic field and the high-energy ($>3$~keV) ions on top of the cold SW beam, or into the SW showing a quiet low magnitude magnetic field and a cold SW beam (without the energetic ion component).
Panel (c) shows the probability output from the classifier, color-coded with blue (SW), black (IF), yellow (MSH), and red (MSP).
At each point in time, the classifier returns the probability that a distribution corresponds to one of these four regions.
For most of the distributions in the time interval shown, the highest probability assigned by the classifier corresponds to the correct region, as is apparent from comparing panels (a-b) with panel (c).
The exception is for a few cases, for example at 16:25:06 UT where the probability of the IF is around $60\%$ and that of the MSH is around $40\%$ while it is clear from panels (a-b) that this is a turbulent MSH region.
Finally, panel (d) shows the situation in which a shock crossing occurs, marked by $1$ or $-1$, depending on if it is a transition from upstream to downstream or vice versa.
These values are multiplied by 2 if the time matches a burst mode interval from MMS.
In total, all 12 shock crossings observed during this interval were correctly identified, suggesting that the method performs well.

The event presented in figure \ref{fig:corrossings} is from November 2018.
We have also performed initial identification of the shock crossings in the entire MMS dataset (starting from 2015) using the presented approach relying on classification by the two CNN models trained on November/December 2017 data.
Our preliminary results are currently highly promising and indicate that rather simple models as the ones presented here can be used to efficiently automate certain tasks on the large MMS dataset.

\section{Summary}
\label{sec:conclusions}
We present a method to automatically detect plasma regions traversed by the MMS spacecraft in its dayside campaign.
This technique is based on the classification of the 3D ion distributions (sky-maps) measured by the FPI DIS instrument using a 3D convolutional neural network (CNN).
The approach is similar to the image recognition using 2D CNNs, but in our case, the images (ion distributions) also have the third dimension corresponding to ion energy.
We have used two different months (November and December 2017) of observations to train and cross-validate two models.
The cross-validations yield the prediction accuracy of $98.4$\% and $98.9$\% for the two models, respectively.
The predictions of the two models are quite similar, and either one can be applied successfully.

Both models are excellent at detecting the solar wind, magnetosheath, and magnetosphere.
The dominant contribution to the model's error is mislabelling of the ion foreshock regions as the solar wind.
However, the model provides low probability scores to such mislabeled examples, and they could be easily identified.
Predictions, where two different classes have rather high probability scores, represent the mixed/boundary regions.
They could be used to detect regions such as the bow shock or magnetopause.

We propose a method based on the obtained probabilities to identify bow shock crossings.
We analyze the time evolution of the probabilities and identify the shock crossings as fast transitions between the magnetosheath and solar wind/foreshock regions.
A similar approach can be used to identify the magnetopause crossings.

\section{Discussion}

The model, proposed in this work, exhibits excellent accuracy on predicting the dayside plasma regions traversed by the MMS. The same model, or a similar one, could be trained to predict plasma environments from the nightside campaign as well. Therefore, it fits well in the concept of the Ground Loop Hierarchy proposed by \citet{Argall:etal:2020}, which aims at full automation of the burst control. The proposed end-to-end approach does not consume any human-engineered features and requires minimal data reduction. Our model has been trained on Level-2 data, constructed from the uncalibrated raw data, calibration data, ancillary, and other data \citep{Baker:2016}. Given the flexibility and adaptability of the convolution neural networks, a similar (perhaps, larger) model could learn to classify the observations given those data, omitting the data reduction step. Some data reduction steps could be represented as layers and embedded in the network architecture itself. Recent developments \citep{Kothari:etal:2020} suggest that neural networks inference can be done directly onboard the satellites, very much like it is done on modern smartphones.
Our model is light-weight, therefore it can be embedded in the software of the future missions, providing an important but cheap telemetry-wise data product, which could be fully transferred to the ground.

We foresee two main directions for further development of our approach.
First, It is useful to compare the predictions of our model for the whole dayside campaign with the SITL reports for the same period. During manual labeling, we have identified the periods which were mislabeled (not well resolved) by the SITL. For instance, intermittent ion foreshock regions in the solar wind data. Not only our model will gain trust in the community in this way, but also it would improve the quality of the models which rely on the SITL labels \citep{Argall:etal:2020,Breuillard:2020}. Second, the model shall be tried against the observations of the nightside campaign. This requires additional manual labeling and re-training of the model on them. Finally, it is interesting to investigate if the same model can be trained to classify both the night- and the day-side plasma regions simultaneously.

Application of our classifier to the existing observations opens new directions for research. 
It would be extremely useful to perform a statistical analysis of the mixed regions detected by the CNN, and figure out whether magnetic reconnection events could be found in those regions.
There is a big potential in studying correlations of different plasma properties measured in different classes, predicted by the classifier.
Large-scale statistical studies of the magnetosphere are often difficult due to the requirement to accurately identify its sub-regions which are bounded by dynamic boundaries (e.g. bow shock and magnetopause), which are sensitively coupled to the solar wind parameters.
Our method offers an efficient way to automatically classify these aforementioned regions, which could facilitate new statistical studies of these regions and boundaries.
We have developed a software package that could be deployed in data centers and be used for automatic classification of observations, statistical analysis, and visualization of the data.

It might be of interest to apply similar techniques to other particle instruments or other spacecraft.
In principle, if different plasma regions exhibit differences in particle distributions, the model should be able to learn these differences and provide accurate classification results.

\acknowledgments
Funding for the work is received from the European Commission H2020 program, Grant 559 Agreement No. 801039 (EPiGRAM-HS).
AD is supported by the Ministry of Science and Higher Education of the Russian Federation under agreement 075-15-2021-583.
APD received financial support from the Swedish National Space Agency (Grant 2020-00111).
The data used in this work is publicly through MMS data archive \url{https://lasp.colorado.edu/mms/sdc/public/about/browse-wrapper/}.
The source codes and the labeled dataset are available on Bitbucket \url{https://bitbucket.org/volshevsky/mmslearning}.


\begin{thebibliography}{}

\bibitem [\protect \citeauthoryear {%
Abadi%
\ \protect \BOthers {.}}{%
Abadi%
\ \protect \BOthers {.}}{%
{\protect \APACyear {2015}}%
}]{%
tensorflow2015-whitepaper}
\APACinsertmetastar {%
tensorflow2015-whitepaper}%
\begin{APACrefauthors}%
Abadi, M.%
, Agarwal, A.%
, Barham, P.%
, Brevdo, E.%
, Chen, Z.%
, Citro, C.%
\BDBL {}Zheng, X.%
\end{APACrefauthors}%
\unskip\
\newblock
\APACrefYearMonthDay{2015}{}{}.
\newblock
\APACrefbtitle {{TensorFlow}: Large-Scale Machine Learning on Heterogeneous
  Systems.} {{TensorFlow}: Large-scale machine learning on heterogeneous
  systems.}
\newblock
\begin{APACrefURL} \url{https://www.tensorflow.org/} \end{APACrefURL}
\newblock
\APACrefnote{Software available from tensorflow.org}
\PrintBackRefs{\CurrentBib}

\bibitem [\protect \citeauthoryear {%
Argall%
\ \protect \BOthers {.}}{%
Argall%
\ \protect \BOthers {.}}{%
{\protect \APACyear {2020}}%
}]{%
Argall:etal:2020}
\APACinsertmetastar {%
Argall:etal:2020}%
\begin{APACrefauthors}%
Argall, M\BPBI R.%
, Small, C\BPBI R.%
, Piatt, S.%
, Breen, L.%
, Petrik, M.%
, Kokkonen, K.%
\BDBL {}Burch, J\BPBI L.%
\end{APACrefauthors}%
\unskip\
\newblock
\APACrefYearMonthDay{2020}{}{}.
\newblock
{\BBOQ}\APACrefatitle {MMS SITL Ground Loop: Automating the Burst Data
  Selection Process} {Mms sitl ground loop: Automating the burst data selection
  process}.{\BBCQ}
\newblock
\APACjournalVolNumPages{Frontiers in Astronomy and Space Sciences}{7}{}{54}.
\newblock
\begin{APACrefURL}
  \url{https://www.frontiersin.org/article/10.3389/fspas.2020.00054}
  \end{APACrefURL}
\newblock
\begin{APACrefDOI} \doi{10.3389/fspas.2020.00054} \end{APACrefDOI}
\PrintBackRefs{\CurrentBib}

\bibitem [\protect \citeauthoryear {%
Baker%
\ \protect \BOthers {.}}{%
Baker%
\ \protect \BOthers {.}}{%
{\protect \APACyear {2016}}%
}]{%
Baker:2016}
\APACinsertmetastar {%
Baker:2016}%
\begin{APACrefauthors}%
Baker, D\BPBI N.%
, Riesberg, L.%
, Pankratz, C\BPBI K.%
, Panneton, R\BPBI S.%
, Giles, B\BPBI L.%
, Wilder, F\BPBI D.%
\BCBL {}\ \BBA {} Ergun, R\BPBI E.%
\end{APACrefauthors}%
\unskip\
\newblock
\APACrefYearMonthDay{2016}{}{}.
\newblock
{\BBOQ}\APACrefatitle {Magnetospheric Multiscale Instrument Suite Operations
  and Data System} {Magnetospheric multiscale instrument suite operations and
  data system}.{\BBCQ}
\newblock
\APACjournalVolNumPages{Space Science Reviews}{199}{}{545-575}.
\newblock
\begin{APACrefDOI} \doi{https://doi.org/10.1007/s11214-014-0128-5}
  \end{APACrefDOI}
\PrintBackRefs{\CurrentBib}

\bibitem [\protect \citeauthoryear {%
Breuillard%
\ \protect \BOthers {.}}{%
Breuillard%
\ \protect \BOthers {.}}{%
{\protect \APACyear {2020}}%
}]{%
Breuillard:2020}
\APACinsertmetastar {%
Breuillard:2020}%
\begin{APACrefauthors}%
Breuillard, H.%
, Dupuis, R.%
, Retino, A.%
, Le~Contel, O.%
, Amaya, J.%
\BCBL {}\ \BBA {} Lapenta, G.%
\end{APACrefauthors}%
\unskip\
\newblock
\APACrefYearMonthDay{2020}{}{}.
\newblock
{\BBOQ}\APACrefatitle {Automatic Classification of Plasma Regions in Near-Earth
  Space With Supervised Machine Learning: Application to Magnetospheric Multi
  Scale 2016�2019 Observations} {Automatic classification of plasma regions
  in near-earth space with supervised machine learning: Application to
  magnetospheric multi scale 2016-2019 observations}.{\BBCQ}
\newblock
\APACjournalVolNumPages{Frontiers in Astronomy and Space Sciences}{7}{}{55}.
\newblock
\begin{APACrefURL}
  \url{https://www.frontiersin.org/article/10.3389/fspas.2020.00055}
  \end{APACrefURL}
\newblock
\begin{APACrefDOI} \doi{10.3389/fspas.2020.00055} \end{APACrefDOI}
\PrintBackRefs{\CurrentBib}

\bibitem [\protect \citeauthoryear {%
Burch%
, Moore%
, Torbert%
\BCBL {}\ \BBA {} Giles%
}{%
Burch%
\ \protect \BOthers {.}}{%
{\protect \APACyear {2016}}%
}]{%
Burch:2016}
\APACinsertmetastar {%
Burch:2016}%
\begin{APACrefauthors}%
Burch, J\BPBI L.%
, Moore, T\BPBI E.%
, Torbert, R\BPBI B.%
\BCBL {}\ \BBA {} Giles, B\BPBI L.%
\end{APACrefauthors}%
\unskip\
\newblock
\APACrefYearMonthDay{2016}{Mar}{01}.
\newblock
{\BBOQ}\APACrefatitle {Magnetospheric Multiscale Overview and Science
  Objectives} {Magnetospheric multiscale overview and science
  objectives}.{\BBCQ}
\newblock
\APACjournalVolNumPages{Space Science Reviews}{199}{1}{5--21}.
\newblock
\begin{APACrefURL} \url{https://doi.org/10.1007/s11214-015-0164-9}
  \end{APACrefURL}
\newblock
\begin{APACrefDOI} \doi{10.1007/s11214-015-0164-9} \end{APACrefDOI}
\PrintBackRefs{\CurrentBib}

\bibitem [\protect \citeauthoryear {%
Chollet%
\ \protect \BOthers {.}}{%
Chollet%
\ \protect \BOthers {.}}{%
{\protect \APACyear {2015}}%
}]{%
chollet2015keras}
\APACinsertmetastar {%
chollet2015keras}%
\begin{APACrefauthors}%
Chollet, F.%
\BCBT {}\ \BOthersPeriod {.}
\end{APACrefauthors}%
\unskip\
\newblock
\APACrefYearMonthDay{2015}{}{}.
\newblock
\APACrefbtitle {Keras.} {Keras.}
\newblock
\APAChowpublished {\url{https://keras.io}}.
\PrintBackRefs{\CurrentBib}

\bibitem [\protect \citeauthoryear {%
Fuselier%
\ \protect \BOthers {.}}{%
Fuselier%
\ \protect \BOthers {.}}{%
{\protect \APACyear {2016}}%
}]{%
Fuselier:2016}
\APACinsertmetastar {%
Fuselier:2016}%
\begin{APACrefauthors}%
Fuselier, S\BPBI A.%
, Lewis, W\BPBI S.%
, Schiff, C.%
, Ergun, R.%
, Burch, J\BPBI L.%
, Petrinec, S\BPBI M.%
\BCBL {}\ \BBA {} Trattner, K\BPBI J.%
\end{APACrefauthors}%
\unskip\
\newblock
\APACrefYearMonthDay{2016}{Mar}{01}.
\newblock
{\BBOQ}\APACrefatitle {Magnetospheric Multiscale Science Mission Profile and
  Operations} {Magnetospheric multiscale science mission profile and
  operations}.{\BBCQ}
\newblock
\APACjournalVolNumPages{Space Science Reviews}{199}{1}{77--103}.
\newblock
\begin{APACrefURL} \url{https://doi.org/10.1007/s11214-014-0087-x}
  \end{APACrefURL}
\newblock
\begin{APACrefDOI} \doi{10.1007/s11214-014-0087-x} \end{APACrefDOI}
\PrintBackRefs{\CurrentBib}

\bibitem [\protect \citeauthoryear {%
Fuselier%
\ \protect \BOthers {.}}{%
Fuselier%
\ \protect \BOthers {.}}{%
{\protect \APACyear {2017}}%
}]{%
Fuselier:2017}
\APACinsertmetastar {%
Fuselier:2017}%
\begin{APACrefauthors}%
Fuselier, S\BPBI A.%
, Vines, S\BPBI K.%
, Burch, J\BPBI L.%
, Petrinec, S\BPBI M.%
, Trattner, K\BPBI J.%
, Cassak, P\BPBI A.%
\BDBL {}Webster, J\BPBI M.%
\end{APACrefauthors}%
\unskip\
\newblock
\APACrefYearMonthDay{2017}{}{}.
\newblock
{\BBOQ}\APACrefatitle {Large-scale characteristics of reconnection diffusion
  regions and associated magnetopause crossings observed by MMS} {Large-scale
  characteristics of reconnection diffusion regions and associated magnetopause
  crossings observed by mms}.{\BBCQ}
\newblock
\APACjournalVolNumPages{Journal of Geophysical Research: Space
  Physics}{122}{5}{5466-5486}.
\newblock
\begin{APACrefURL}
  \url{https://agupubs.onlinelibrary.wiley.com/doi/abs/10.1002/2017JA024024}
  \end{APACrefURL}
\newblock
\begin{APACrefDOI} \doi{10.1002/2017JA024024} \end{APACrefDOI}
\PrintBackRefs{\CurrentBib}

\bibitem [\protect \citeauthoryear {%
Hotelling%
}{%
Hotelling%
}{%
{\protect \APACyear {1933}}%
}]{%
Hotelling:1933}
\APACinsertmetastar {%
Hotelling:1933}%
\begin{APACrefauthors}%
Hotelling, H.%
\end{APACrefauthors}%
\unskip\
\newblock
\APACrefYearMonthDay{1933}{}{}.
\newblock
{\BBOQ}\APACrefatitle {Analysis of a Complex of Statistical Variables Into
  Principal Components} {Analysis of a complex of statistical variables into
  principal components}.{\BBCQ}
\newblock
\APACjournalVolNumPages{Journal of Educational
  Psychology}{24}{}{417-441,498-520}.
\PrintBackRefs{\CurrentBib}

\bibitem [\protect \citeauthoryear {%
Jolliffe%
}{%
Jolliffe%
}{%
{\protect \APACyear {2002}}%
}]{%
Jolliffe:2002:PCA}
\APACinsertmetastar {%
Jolliffe:2002:PCA}%
\begin{APACrefauthors}%
Jolliffe, I.%
\end{APACrefauthors}%
\unskip\
\newblock
\APACrefYear{2002}.
\newblock
\APACrefbtitle {Principal Component Analysis} {Principal component analysis}.
\newblock
\APACaddressPublisher{}{Springer}.
\newblock
\begin{APACrefURL} \url{https://books.google.se/books?id=\_olByCrhjwIC}
  \end{APACrefURL}
\PrintBackRefs{\CurrentBib}

\bibitem [\protect \citeauthoryear {%
Kingma%
\ \BBA {} Ba%
}{%
Kingma%
\ \BBA {} Ba%
}{%
{\protect \APACyear {2014}}%
}]{%
kingma:2014}
\APACinsertmetastar {%
kingma:2014}%
\begin{APACrefauthors}%
Kingma, D\BPBI P.%
\BCBT {}\ \BBA {} Ba, J.%
\end{APACrefauthors}%
\unskip\
\newblock
\APACrefYearMonthDay{2014}{}{}.
\newblock
{\BBOQ}\APACrefatitle {Adam: A method for stochastic optimization} {Adam: A
  method for stochastic optimization}.{\BBCQ}
\newblock
\APACjournalVolNumPages{arXiv preprint arXiv:1412.6980}{}{}{}.
\PrintBackRefs{\CurrentBib}

\bibitem [\protect \citeauthoryear {%
Kothari%
, Liberis%
\BCBL {}\ \BBA {} Lane%
}{%
Kothari%
\ \protect \BOthers {.}}{%
{\protect \APACyear {2020}}%
}]{%
Kothari:etal:2020}
\APACinsertmetastar {%
Kothari:etal:2020}%
\begin{APACrefauthors}%
Kothari, V.%
, Liberis, E.%
\BCBL {}\ \BBA {} Lane, N\BPBI D.%
\end{APACrefauthors}%
\unskip\
\newblock
\APACrefYearMonthDay{2020}{}{}.
\newblock
{\BBOQ}\APACrefatitle {The Final Frontier: Deep Learning in Space} {The final
  frontier: Deep learning in space}.{\BBCQ}
\newblock
\BIn{} \APACrefbtitle {Proceedings of the 21st International Workshop on Mobile
  Computing Systems and Applications} {Proceedings of the 21st international
  workshop on mobile computing systems and applications}\ (\BPG~45-49).
\newblock
\APACaddressPublisher{New York, NY, USA}{Association for Computing Machinery}.
\newblock
\begin{APACrefURL} \url{https://doi.org/10.1145/3376897.3377864}
  \end{APACrefURL}
\newblock
\begin{APACrefDOI} \doi{10.1145/3376897.3377864} \end{APACrefDOI}
\PrintBackRefs{\CurrentBib}

\bibitem [\protect \citeauthoryear {%
{LeCun}%
, {Bengio}%
\BCBL {}\ \BBA {} {Hinton}%
}{%
{LeCun}%
\ \protect \BOthers {.}}{%
{\protect \APACyear {2015}}%
}]{%
Lecun:Yann:Bengio:2015}
\APACinsertmetastar {%
Lecun:Yann:Bengio:2015}%
\begin{APACrefauthors}%
{LeCun}, Y.%
, {Bengio}, Y.%
\BCBL {}\ \BBA {} {Hinton}, G.%
\end{APACrefauthors}%
\unskip\
\newblock
\APACrefYearMonthDay{2015}{May}{}.
\newblock
{\BBOQ}\APACrefatitle {{Deep learning}} {{Deep learning}}.{\BBCQ}
\newblock
\APACjournalVolNumPages{Nature}{521}{7553}{436-444}.
\newblock
\begin{APACrefDOI} \doi{10.1038/nature14539} \end{APACrefDOI}
\PrintBackRefs{\CurrentBib}

\bibitem [\protect \citeauthoryear {%
Liu%
}{%
Liu%
}{%
{\protect \APACyear {2021}}%
}]{%
nasa-cdf}
\APACinsertmetastar {%
nasa-cdf}%
\begin{APACrefauthors}%
Liu, M.%
\end{APACrefauthors}%
\unskip\
\newblock
\APACrefYearMonthDay{2021}{}{}.
\newblock
\APACrefbtitle {Common Data Format (CDF).} {Common data format (cdf).}
\newblock
\APAChowpublished {\url{https://cdf.gsfc.nasa.gov}}.
\newblock
\APACrefnote{[Online; accessed 13-August-2021]}
\PrintBackRefs{\CurrentBib}

\bibitem [\protect \citeauthoryear {%
{Maturana}%
\ \BBA {} {Scherer}%
}{%
{Maturana}%
\ \BBA {} {Scherer}%
}{%
{\protect \APACyear {2015}}%
}]{%
Maturana:Scherer:2015}
\APACinsertmetastar {%
Maturana:Scherer:2015}%
\begin{APACrefauthors}%
{Maturana}, D.%
\BCBT {}\ \BBA {} {Scherer}, S.%
\end{APACrefauthors}%
\unskip\
\newblock
\APACrefYearMonthDay{2015}{Sep.}{}.
\newblock
{\BBOQ}\APACrefatitle {VoxNet: A 3D Convolutional Neural Network for real-time
  object recognition} {Voxnet: A 3d convolutional neural network for real-time
  object recognition}.{\BBCQ}
\newblock
\BIn{} \APACrefbtitle {2015 IEEE/RSJ International Conference on Intelligent
  Robots and Systems (IROS)} {2015 ieee/rsj international conference on
  intelligent robots and systems (iros)}\ (\BPG~922-928).
\newblock
\begin{APACrefDOI} \doi{10.1109/IROS.2015.7353481} \end{APACrefDOI}
\PrintBackRefs{\CurrentBib}

\bibitem [\protect \citeauthoryear {%
{MAVEN SDC}%
}{%
{MAVEN SDC}%
}{%
{\protect \APACyear {2021}}%
}]{%
cdflib}
\APACinsertmetastar {%
cdflib}%
\begin{APACrefauthors}%
{MAVEN SDC}.%
\end{APACrefauthors}%
\unskip\
\newblock
\APACrefYearMonthDay{2021}{}{}.
\newblock
\APACrefbtitle {CDFLib.} {Cdflib.}
\newblock
\APAChowpublished {\url{https://github.com/MAVENSDC/cdflib}}.
\newblock
\APACrefnote{[Online; accessed 13-August-2021]}
\PrintBackRefs{\CurrentBib}

\bibitem [\protect \citeauthoryear {%
{MMS Science Data Center}%
}{%
{MMS Science Data Center}%
}{%
{\protect \APACyear {2021}}%
}]{%
mms-sdc}
\APACinsertmetastar {%
mms-sdc}%
\begin{APACrefauthors}%
{MMS Science Data Center}.%
\end{APACrefauthors}%
\unskip\
\newblock
\APACrefYearMonthDay{2021}{}{}.
\newblock
\APACrefbtitle {MMS Science Data Center.} {Mms science data center.}
\newblock
\APAChowpublished {\url{https://lasp.colorado.edu/mms/sdc/public/}}.
\newblock
\APACrefnote{[Online; accessed 13-August-2021]}
\PrintBackRefs{\CurrentBib}

\bibitem [\protect \citeauthoryear {%
Olshevsky%
}{%
Olshevsky%
}{%
{\protect \APACyear {2021}}%
}]{%
mmslearning}
\APACinsertmetastar {%
mmslearning}%
\begin{APACrefauthors}%
Olshevsky, V.%
\end{APACrefauthors}%
\unskip\
\newblock
\APACrefYearMonthDay{2021}{}{}.
\newblock
\APACrefbtitle {MMSLearning.} {Mmslearning.}
\newblock
\APAChowpublished {\url{https://bitbucket.org/volshevsky/mmslearning}}.
\newblock
\APACrefnote{[Online; accessed 13-August-2021]}
\PrintBackRefs{\CurrentBib}

\bibitem [\protect \citeauthoryear {%
Pearson%
}{%
Pearson%
}{%
{\protect \APACyear {1901}}%
}]{%
Pearson:1901:PCA}
\APACinsertmetastar {%
Pearson:1901:PCA}%
\begin{APACrefauthors}%
Pearson, K.%
\end{APACrefauthors}%
\unskip\
\newblock
\APACrefYearMonthDay{1901}{}{}.
\newblock
{\BBOQ}\APACrefatitle {LIII. On lines and planes of closest fit to systems of
  points in space} {Liii. on lines and planes of closest fit to systems of
  points in space}.{\BBCQ}
\newblock
\APACjournalVolNumPages{The London, Edinburgh, and Dublin Philosophical
  Magazine and Journal of Science}{2}{11}{559-572}.
\newblock
\begin{APACrefURL} \url{https://doi.org/10.1080/14786440109462720}
  \end{APACrefURL}
\newblock
\begin{APACrefDOI} \doi{10.1080/14786440109462720} \end{APACrefDOI}
\PrintBackRefs{\CurrentBib}

\bibitem [\protect \citeauthoryear {%
Pedregosa%
\ \protect \BOthers {.}}{%
Pedregosa%
\ \protect \BOthers {.}}{%
{\protect \APACyear {2011}}%
}]{%
scikit-learn}
\APACinsertmetastar {%
scikit-learn}%
\begin{APACrefauthors}%
Pedregosa, F.%
, Varoquaux, G.%
, Gramfort, A.%
, Michel, V.%
, Thirion, B.%
, Grisel, O.%
\BDBL {}Duchesnay, E.%
\end{APACrefauthors}%
\unskip\
\newblock
\APACrefYearMonthDay{2011}{}{}.
\newblock
{\BBOQ}\APACrefatitle {Scikit-learn: Machine Learning in {P}ython}
  {Scikit-learn: Machine learning in {P}ython}.{\BBCQ}
\newblock
\APACjournalVolNumPages{Journal of Machine Learning
  Research}{12}{}{2825--2830}.
\PrintBackRefs{\CurrentBib}

\bibitem [\protect \citeauthoryear {%
{Piatt}%
}{%
{Piatt}%
}{%
{\protect \APACyear {2019}}%
}]{%
Piatt:2019arXiv}
\APACinsertmetastar {%
Piatt:2019arXiv}%
\begin{APACrefauthors}%
{Piatt}, S.%
\end{APACrefauthors}%
\unskip\
\newblock
\APACrefYearMonthDay{2019}{May}{}.
\newblock
{\BBOQ}\APACrefatitle {{Large-Scale Statistical Survey of Magnetopause
  Reconnection}} {{Large-Scale Statistical Survey of Magnetopause
  Reconnection}}.{\BBCQ}
\newblock
\APACjournalVolNumPages{arXiv e-prints}{}{}{arXiv:1905.11359}.
\PrintBackRefs{\CurrentBib}

\bibitem [\protect \citeauthoryear {%
Pollock%
\ \protect \BOthers {.}}{%
Pollock%
\ \protect \BOthers {.}}{%
{\protect \APACyear {2016}}%
}]{%
Pollock:2016}
\APACinsertmetastar {%
Pollock:2016}%
\begin{APACrefauthors}%
Pollock, C.%
, Moore, T.%
, Jacques, A.%
, Burch, J.%
, Gliese, U.%
, Saito, Y.%
\BDBL {}Zeuch, M.%
\end{APACrefauthors}%
\unskip\
\newblock
\APACrefYearMonthDay{2016}{Mar}{01}.
\newblock
{\BBOQ}\APACrefatitle {Fast Plasma Investigation for Magnetospheric Multiscale}
  {Fast plasma investigation for magnetospheric multiscale}.{\BBCQ}
\newblock
\APACjournalVolNumPages{Space Science Reviews}{199}{1}{331--406}.
\newblock
\begin{APACrefURL} \url{https://doi.org/10.1007/s11214-016-0245-4}
  \end{APACrefURL}
\newblock
\begin{APACrefDOI} \doi{10.1007/s11214-016-0245-4} \end{APACrefDOI}
\PrintBackRefs{\CurrentBib}

\bibitem [\protect \citeauthoryear {%
Russell%
\ \protect \BOthers {.}}{%
Russell%
\ \protect \BOthers {.}}{%
{\protect \APACyear {2016}}%
}]{%
Russell:2016}
\APACinsertmetastar {%
Russell:2016}%
\begin{APACrefauthors}%
Russell, C\BPBI T.%
, Anderson, B\BPBI J.%
, Baumjohann, W.%
, Bromund, K\BPBI R.%
, Dearborn, D.%
, Fischer, D.%
\BDBL {}Richter, I.%
\end{APACrefauthors}%
\unskip\
\newblock
\APACrefYearMonthDay{2016}{Mar}{01}.
\newblock
{\BBOQ}\APACrefatitle {The Magnetospheric Multiscale Magnetometers} {The
  magnetospheric multiscale magnetometers}.{\BBCQ}
\newblock
\APACjournalVolNumPages{Space Science Reviews}{199}{1}{189--256}.
\newblock
\begin{APACrefURL} \url{https://doi.org/10.1007/s11214-014-0057-3}
  \end{APACrefURL}
\newblock
\begin{APACrefDOI} \doi{10.1007/s11214-014-0057-3} \end{APACrefDOI}
\PrintBackRefs{\CurrentBib}

\end{thebibliography}
\end{document}